\documentclass[prc,superscriptaddress,unsortedaddress,
twocolumn,showpacs,preprintnumbers,amsmath,amssymb]{revtex4}
\usepackage[dvips]{graphicx}
\usepackage{dcolumn}
\usepackage{amsmath}
\usepackage{times}
\def\la{\mathrel{\mathpalette\fun <}}
\def\ga{\mathrel{\mathpalette\fun >}}
\def\fun#1#2{\lower3.6pt\vbox{\baselineskip0pt\lineskip.9pt
  \ialign{$\mathsurround=0pt#1\hfil##\hfil$\crcr#2\crcr\sim\crcr}}}

\begin{document}

\title{ Determination of $S_{17}$ from ${}^8$B breakup by means
of the method of continuum-discretized coupled-channels}

\author{K. Ogata}
\email[Electronic address: ]{kazu2scp@mbox.nc.kyushu-u.ac.jp}
\affiliation{Department of Physics, Kyushu University,
Fukuoka 812-8581,
Japan}
\author{S. Hashimoto}
\affiliation{Department of Physics, Kyushu University,
Fukuoka 812-8581,
Japan}
\author{Y. Iseri}
\affiliation{Department of Physics, Chiba-Keizai College,
Todoroki-cho
4-3-30, Inage, Chiba 263-0021, Japan}
\author{M. Kamimura}
\affiliation{Department of Physics, Kyushu University,
Fukuoka 812-8581,
Japan}
\author{M. Yahiro}
\affiliation{Department of Physics, Kyushu University,
Fukuoka 812-8581,
Japan}

\date{\today}

\begin{abstract}
The astrophysical factor for $^7$Be($p,\gamma$)$^8$B
at zero energy, $S_{17}(0)$, is determined from an analysis of
$^{208}$Pb($^8$B, $p+^7$Be)$^{208}$Pb at 52 MeV/nucleon
by means of the method of continuum-discretized coupled-channels
(CDCC)
taking account of all nuclear and Coulomb breakup
processes.
The asymptotic normalization coefficient (ANC)
method is used to extract $S_{17}(0)$ from the calculated
breakup cross section. The main result of the present paper is
$S_{17}(0)=20.9^{+2.0}_{-1.9}$ eV b.
The error consists of 8.4\% experimental systematic error and
the error due to the ambiguity in the s-wave $p$-$^7$Be scattering length.
This value of $S_{17}(0)$ differs from the one extracted with the
first-order perturbation theory including Coulomb breakup by
dipole transitions: $18.9 \pm 1.8$ eV b.
It turns out that the difference is due to the
inclusion of the nuclear and Coulomb-quadrupole transitions and multi-step
processes of all-order in the present work.
The p-$^7$Be interaction potential used in the CDCC calculation
is also used in the ANC analysis of $^7$Be($p,\gamma$)$^8$B.
The value of $S_{17}(0)=21.7^{+0.62}_{-0.55}$ eV b obtained is
consistent with the previous one
obtained from a precise measurement
of the $p$-capture reaction cross section extrapolated to
zero incident energy,
$S_{17}(0)=22.1 \pm 0.6$ (expt) $\pm 0.6$ (theo) eV b, where (theo)
stands for the error in the extrapolation.
Thus, the agreement between
the values of $S_{17}(0)$ obtained from direct
$^7$Be($p,\gamma$)$^8$B and indirect $^8$B-breakup measurements
is significantly improved.
\end{abstract}


\maketitle

\section{Introduction}
\label{intro}

The solar neutrino problem is one of the central issues in
the neutrino physics~\cite{Bahcall}.
The current interpretation of the solar-neutrino deficit is
the neutrino oscillation induced by
the mass difference among $\nu_{e}$, $\nu_{\mu}$, and $\nu_{\tau}$,
and their mixing angles~\cite{Bahcall2}.
Attention is now focused on
the determination of the oscillation parameters.
The cross section of the $p$-capture reaction $^7$Be($p,\gamma$)$^8$B
at incident energy in the center-of-mass (c.m.) frame of the
$p$-$^7$Be system $E_{\rm c.m.} \sim 0$
plays an essential role in the solar-neutrino phenomenology,
since the observed flux of $^8$B neutrino is proportional to it;
the magnitude of the cross section $\sigma_{p\gamma}(E_{\rm c.m.})$
is customarily
expressed by the astrophysical factor
$S_{17}(E_{\rm c.m.})
\equiv \sigma_{p\gamma}(E_{\rm c.m.})E_{\rm c.m.}\exp [2\pi \eta]$,
where $\eta$ is the Sommerfeld parameter.
The required accuracy of $S_{17}(0)$, to determine the neutrino
oscillation parameters with sufficient accuracy, is the error
within 5\%~\cite{Bahcall3}.

Recently, precise measurement of $\sigma_{p\gamma}(E_{\rm c.m.})$
was carried out
by Junghans {\it et al.}~\cite{Junghans2}
at energies of $E_{\rm c.m.}=$116--2460 keV, which are low
but still higher than stellar energies ($\sim 20$ keV).
Extrapolating the measured $S_{17}(E_{\rm c.m.})$
to $E_{\rm c.m.} = 0$
using a three-cluster model~\cite{DB} for $^8$B structure,
they derived $S_{17}(0)=22.1 \pm 0.6$ (expt) $\pm 0.6$ (theo) eV b.
The three-body model, however, did not simultaneously reproduce the
magnitude and the energy-dependence of $S_{17}(E_{\rm c.m.})$
sufficiently well.
Moreover, as pointed out in Ref.~\cite{DB2}, the uncertainty of
the s-wave $p$-$^7$Be scattering length (with about 50\% error)
prevents one from determining $S_{17}(0)$ with very high accuracy.

Because of the difficulty of the direct measurement of
$\sigma_{p\gamma}(E_{\rm c.m.})$ at stellar energies,
alternative indirect measurements were proposed. Coulomb
dissociation~\cite{RIKEN1,RIKEN2,RIKEN3,GSI1,GSI2,MSU1,MSU2} of $^8$B
is one of such indirect measurements.
So far, extraction of $S_{17}(0)$ from these experiments has been
based on the virtual photon theory with the assumption of
$^8$B dissociation by virtual electric dipole (E1) photon absorption.
Nuclear interaction and absorption of quadrupole (E2)- and
multi-photons were not taken into account.
The value thus extracted from the RIKEN experiment at 52 MeV/nucleon
was $S_{17}(0)=18.9 \pm 1.8$ eV b~\cite{RIKEN3}.
In the analysis of the MSU experiment~\cite{MSU1,MSU2} measured at
44 and 83 MeV/nucleon,
the E2 contribution to one-step process was
estimated from the parallel-momentum-distribution
of $^7$Be fragment and subtracted from the breakup spectrum of $^8$B.
As a result, the extracted $S_{17}(0)$ was $17.8^{ +1.4}_{-1.2}$ eV b,
which is smaller than the value obtained at RIKEN mentioned above.
However, the analysis of the angular distribution of the $^8$B
breakup cross section of the RIKEN experiment
showed no contribution of E2 transitions.
Moreover, the recent experiment at rather high energy, 250 MeV/nucleon,
done at GSI~\cite{GSI2} showed that the E2 contribution was negligibly
small; the resulting $S_{17}(0)$ was $18.6 \pm 1.2 \pm 1.0$ eV b.
Even though the significance of E2 transitions can be energy-dependent,
the conclusions from the MSU and RIKEN measurements
at similar energies look
inconsistent and roles of the E2 component are still not clear.
More seriously, there exists a non-negligible discrepancy of
about 15\% between the values of
$S_{17}(0)$ mentioned above that are derived from direct $p$-capture
and indirect $^8$B-breakup measurements.

Very recently, it was shown by a semiclassical calculation
of $^{208}$Pb($^8$B, $p+^7$Be)$^{208}$Pb at 52 MeV/nucleon
that the discrepancy mentioned above was significantly reduced
by taking account of nuclear-breakup components, E2, and
higher-order Coulomb breakup processes~\cite{EB-PRL}.
Motivated by this result,
we attempt in the present paper
to extract a reliable value of $S_{17}(0)$
analyzing the $^8$B dissociation experiment measured at
RIKEN~\cite{RIKEN1,RIKEN2,RIKEN3}, with
the method of continuum-discretized
coupled-channels~\cite{CDCC-review1,CDCC-review2} (CDCC),
assuming a $p$+${^7}$Be+A three-body model of the system,
where A stands for the target nucleus.
The result of the analysis is then used to extract $S_{17}(0)$
by means of the asymptotic normalization coefficient (ANC)
method~\cite{Akram}.
CDCC has been successful in describing
various processes in which effects of projectile-breakup are
essential~\cite{Yahiro1,yahiro2,sakuragi1,%
sakuragi2,sakuragi34,iseri1,Sakuragi,Surrey,Ogata}.
It has been successfully applied also to $^8$B nuclear and
Coulomb breakup processes~\cite{MSU2,Tostevin,Mortimer,OgataND}.
In the CDCC calculation of the present work,
we include all nuclear and Coulomb breakup processes, and take account of
the intrinsic spins of both $p$ and $^7$Be using the channel spin
representation and the consistency of the $p$-$^7$Be interaction
potential used in the CDCC calculation with the s-wave $p$-$^7$Be
scattering length.
The wave function of $^8$B used in the CDCC calculation is
found to be consistent with the measured cross section of
$^7$Be($p,\gamma$)$^8$B at low energies.
We then use the calculated breakup cross section and the $p$-$^7$Be
separation energy in the $^8$B nucleus
to obtain $S_{17}(0)$ by the ANC method.
An important advantage of the ANC method
is that there is no restriction of the reaction mechanisms.
In addition, the uncertainty of $S_{17}(0)$ due to the use of the ANC
method can quantitatively be
evaluated~\cite{Ogata,OgataND}.

The construction of the present paper is as follows.
First, we describe how to extract the ANC from the three-body model
CDCC analysis of $^8$B breakup.
We show that under a condition, which is satisfied in the present
analysis, the calculated cross section is proportional to the
square of the ANC for the ground state of $^8$B.
This is by no means trivial since the CDCC wave function contains
not only the ground state of $^8$B but also its continuum states.
We describe in the later section the condition and the angular
region of $^8$B breakup in which it is satisfied.
Second, we determine $S_{17}(0)$
from $^{208}$Pb($^8$B, $p+^7$Be)$^{208}$Pb at 52 MeV/nucleon by
the ANC method, taking account of all order nuclear and Coulomb
breakup processes.
It should be noted that in our CDCC calculation
interference between nuclear- and Coulomb-breakup amplitudes is
explicitly included.
We make use of
the eikonal-CDCC method~\cite{Ogata2} (E-CDCC), with some modifications,
to obtain the scattering amplitudes corresponding to large values of the
orbital-angular-momentum $L$ of relative motion
between the projectile and the target nucleus.
We then construct a hybrid amplitude from the results of E-CDCC
and CDCC.
This makes it possible to carry out very
accurate analyses with high computational speed~\cite{Ogata2}.
Uncertainties of the extracted $S_{17}(0)$
are carefully evaluated.
Finally, we discuss the contributions of nuclear breakup,
E2 transitions, and higher-order processes
on the extracted value of $S_{17}(0)$.
The ANC method is also applied to direct
$^7$Be($p,\gamma$)$^8$B measurement assuming the same $p$-$^7$Be potential
in the calculation of the direct capture process as used in
the CDCC analysis.
We show the difference between the values of $S_{17}(0)$ derived from
direct and indirect measurements is essentially removed by the
present analysis.

In Sec.~\ref{ANC} we describe how to extract the ANC from the
CDCC analysis of $^8$B breakup.
Formulation of modified E-CDCC and construction of the hybrid scattering
amplitude are described in Sec.~\ref{E-CDCC}.
We analyze in Sec.~\ref{resS17}
the $^8$B breakup cross section measured at
RIKEN with CDCC and extract $S_{17}(0)$ by the ANC method.
Uncertainties of the extracted $S_{17}(0)$
are quantitatively evaluated in Sec.~\ref{errorS17}.
In Sec.~\ref{discussion}, the value of the extracted $S_{17}(0)$ is
compared with the result of the virtual photon theory
and the roles of nuclear breakup,
E2 transitions, and higher-order processes are discussed.
We discuss in Sec.~\ref{direct} the application of the ANC method to
precise direct measurement of $^7$Be($p,\gamma$)$^8$B that confirms
the consistency of its result with the $^8$B breakup analysis.
Finally summary and conclusions are given in Sec.~\ref{summary}.

\section{Formulation}

\subsection{Extraction of ANC from $^8$B breakup}
\label{ANC}

In this subsection we describe our method of extracting $S_{17}(0)$
by $p$+$^7$Be+A three-body model CDCC analysis of elastic $^8$B breakup
reaction combined with the ANC method~\cite{Akram}.
We recapitulate the ANC method below,
partly to introduce notations.
At very low incident energies,
the $p$-capture reaction $^7$Be($p,\gamma$)$^8$B
leading to the ground state of $^8$B
is extremely peripheral because of the Coulomb barrier.
The $T$-matrix element of the reaction, therefore, depends on
the overlap between the ground state wave functions of
$^7$Be and $^8$B only in the tail region,
where the radial part of the overlap has the following form
\begin{eqnarray}
\varphi_{\rm r}^{(0)}(r)
       &\equiv&
       \langle
        \phi^{(0)}_{7} (\xi_7)| \phi^{(0)}_{8}(\xi_7,{\bf r})
       \rangle_{\xi_7, \hat{\bf r}} \nonumber \\
       &\sim&
       C W_{-\eta, \ell_0+1/2}(2k_0 r) / r, \quad
    r > R_{\rm N},
\label{8Bwf}
\end{eqnarray}
where
$R_{\rm N}$ is the range of the nuclear interaction between $p$ and
$^7$Be of about 4 fm,
$\phi^{(0)}_{7}$ ($\phi^{(0)}_{8}$) is the ground state wave function of
$^7$Be ($^8$B),
$k_0$ is the relative momentum
in the unit of $\hbar$
between $p$ and $^7$Be,
$\xi_7$ stands for the internal coordinates of the seven-nucleon
system, and
$W_{-\eta, \ell_0+1/2}$
is the Whittaker function where
$\ell_0$ is the orbital angular momentum
of the $p$-$^7$Be relative motion
in the ground state of $^8$B; $\ell_0=1$ in the present case.
The constant $C$ in Eq.~(\ref{8Bwf}) is the ANC of our interest.
The cross section $\sigma_{p \gamma}(0)$, and consequently $S_{17}(0)$, are
proportional to $C^2$. Thus, one can evaluate them once the ANC, $C$,
is known.
The cross section of any reaction in which a proton
is transferred to or from $^8$B is proportional to $C^2$ in DWBA
if the reaction is peripheral in the
sense that it takes place when the transferred proton is distant from
$^7$Be more than the range $R_{\rm N}$. Based on this idea, DWBA analyses
of
proton transfer reactions
such as $^{10}$Be($^7$Be,$^8$B)$^{9}$Be~\cite{Azhari1}
and $^{14}$N($^7$Be,$^8$B)$^{13}$C~\cite{Azhari2}
have successfully been made to
obtain the required ANC.

In the present paper we analyze the data of $^8$B elastic breakup in
which the fragments $^7$Be and $p$ are ejected into very forward angles.
For brevity, we henceforth call such a process breakup into forward angles.
Since the breakup in such a case takes place at large distance $R$ of
$^8$B from the target nucleus, it is mostly due to Coulomb interaction.
The operator of the interaction is proportional to $r^n$ with $n \ge 1$.
Therefore, the dominant contribution to the $T$ matrix is from the
region of $r$ larger than the range of nuclear interaction $R_{\rm N}$
between the fragments. This is numerically confirmed as described in
Sec.~\ref{errorS17}.
The reaction, therefore, is peripheral in the sense described
above.
Thus, the ANC can be obtained if its cross section is reproduced and is
shown to be proportional to $C^2$ by theory. We show in the following that
the cross section is well reproduced
by CDCC and is in good approximation
proportional to $C^2$ in the present case.

We describe $^8$B breakup at 52 MeV/nucleon with the
$p$+$^7$Be+A three-body model illustrated in Fig.~1.
%
\begin{figure}[htbp]
\begin{center}
 \includegraphics[width=0.35\textwidth,clip]{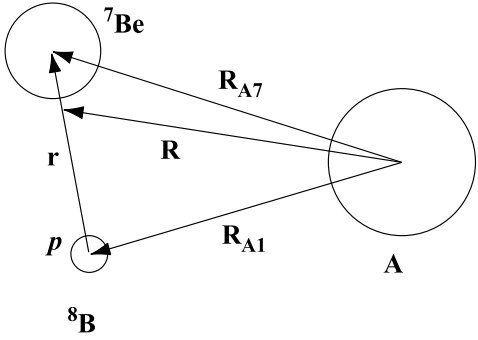}
 \caption{Illustration of the $p$+$^7$Be+A three-body system,
 where A is a target.}
\end{center}
\end{figure}
Let us assume for simplicity that A is infinitely heavy.
The triple differential cross section
for ($^8$B,$^7$Be$+p$) reaction
is given by ${\cal D} \rho |{\mathfrak{T}}|^2$
with a given constant ${\cal D}$
and the three-body phase space factor $\rho$~\cite{Iseri}.
The $T$-matrix element of $^8$B breakup is
\begin{equation}
{\mathfrak{T}}
=
\langle \chi_{1}
\chi_{7}
\phi^{(0)}_{7}|
V
|\Psi \rangle,
\label{T}
\end{equation}
with
\[
V
=
v_{\rm A7}({\bf R}_{\rm A7},\xi_7)
+U_{\rm A1}({\bf R}_{\rm A1})
+v_{\rm 17}({\bf r},\xi_7),
\]
where
$v_{\rm XY}$ ($U_{\rm XY}$)
is the interaction (distorting potential)
between two nuclei X and Y, where the
subscripts 1, 7, and A stand for $p$, $^7$Be, and A, respectively.
The function $\chi_{1}$ ($\chi_{7}$) is the plane wave of
relative motion between A and the outgoing $p$ ($^7$Be),
and $\Psi$ is the exact wave function of the system
with the incident plane wave of relative motion between
$^8$B and A in the initial channel.

In the RIKEN experiment that we analyze in the present work,
emitted fragments were measured at forward angles with the $^7$Be
in the ground state~\cite{RIKEN1,RIKEN2,RIKEN3}.
We assume that $^7$Be stays in its ground state in the course of $^8$B
breakup.
We have confirmed the validity of this assumption
by a $^3$He+$^4$He+$^{208}$Pb
three-body model CDCC calculation
of the scattering of $^7$Be on $^{208}$Pb at
52 MeV/nucleon.
The cross sections of non-elastic processes including breakup of $^7$Be
are less than about 1\% of that of elastic scattering
for the scattering angle of the c.m. of $^7$Be,
$\theta_7$, below about $4^\circ$.
In general $\theta_7$ can differ from the
$^8$B scattering angle $\theta_8$ in the
$^8$B$ + ^{208}$Pb $\rightarrow p+^7$Be $+ ^{208}$Pb
process,
but the difference between the two angles is very small,
since the energy transfer in the $^8$B breakup concerned
is much smaller than the incident energy of $^8$B.
These facts imply that in ${\mathfrak{T}}$ one can neglect matrix elements
of $v_{\rm A7}$ between the ground state and excited states
of $^7$Be. One can, therefore, approximate $v_{\rm A7}$
by the single-folding potential
\[
U_{\rm {A7}}({\bf R}_{\rm A7})
\equiv
\langle
\phi^{(0)}_{7}(\xi_7)
|
v_{\rm A7}({\bf R}_{\rm A7},\xi_7)
|
\phi^{(0)}_{7}(\xi_7)
\rangle_{\xi_7}.
\]
Similarly,
$v_{\rm 17}$ can be approximated by
\[
V_{\rm {17}}({\bf r})
\equiv
\langle
\phi^{(0)}_{7}(\xi_7)
|
v_{\rm 17}({\bf r},\xi_7)
|\phi^{(0)}_{7}(\xi_7)
\rangle_{\xi_7}.
\]
In this approximation,
${\mathfrak{T}}$ is given by
\begin{equation}
{\mathfrak{T}}
\approx \langle \chi_{1}\chi_{7}|
U_{\rm {A7}} + U_{\rm {A1}} + V_{\rm {17}}|
{\cal O} \rangle
\label{Tap0}
\end{equation}
with
\begin{equation}
{\cal O}
\equiv
\langle\phi^{(0)}_{7}|\Psi \rangle.
\end{equation}
If the incident momentum of the $^8$B projectile is ${\bf P}$,
one can make use of the well-known formula for
$\Psi$ and write
\begin{equation}
{\cal O}
=
\Big\langle
 \phi^{(0)}_{7}\Big|
   \frac{i\varepsilon}{E-H+i\varepsilon}
 \Big|\phi^{(0)}_{8}e^{i{\bf P} \cdot {\bf R}}
\Big\rangle,
\label{omega}
\end{equation}
where
$H$ is the exact Hamiltonian of the system
\[
H = T_{r} + T_{R} + V +h_7,
\]
where
$T_{r}$ and $T_{R}$ are the kinetic energy operators
with respect to ${\bf r}$ and ${\bf R}$, respectively,
and $h_7$ is the internal Hamiltonian of $^7$Be.
The total energy of the system is denoted by
$E$, which is related to the
c.m. incident energy of $^8$B $E_{\rm in}$ by
$E=E_{\rm in}+e_7+e_{17}$,
where $e_{7}$ is the eigenenergy of $\phi^{(0)}_{7}$
and $e_{17}$ is the relative energy between $p$ and $^7$Be
in the ground state of $^8$B.
Under the assumption of the inert $^7$Be nucleus mentioned above,
one can approximate $H$ in Eq.~(\ref{omega}) by
\begin{equation}
H \approx e_{7} + H_{\rm 3\mbox{-}body},
\label{H4toH3}
\end{equation}
where
$H_{\rm 3\mbox{-}body}$ is the $p$+$^7$Be+A
three-body Hamiltonian defined by
\begin{equation}
H_{\rm 3\mbox{-}body} = T_{r} + T_{R} +  V_{\rm 17}({\bf r})
   + U_{\rm A7}({\bf R}_{\rm {A7}}) + U_{\rm A1}({\bf R}_{\rm {A1}}).
\label{H}
\end{equation}
Consequently,
\begin{equation}
{\mathfrak{T}}
\approx
\Big\langle \chi_{1}\chi_{7}\Big|U_{\rm {A7}} +
U_{\rm {A1}} + V_{\rm {17}}\Big|
\Omega^{(+)}
\varphi^{(0)}({\bf r})e^{i{\bf P} \cdot {\bf R}}
\Big\rangle,
\label{Tap}
\end{equation}
where
\begin{equation}
\Omega^{(+)}
\equiv
\frac{i\varepsilon}{E- e_{7} -H_{\rm 3\mbox{-}body}+i\varepsilon}
\label{wavematrix}
\end{equation}
is
M$\phi$ller's
wave matrix with the
outgoing boundary condition
and
\begin{equation}
\varphi^{(0)} ({\bf r})\equiv
\langle \phi^{(0)}_{7} (\xi_7)| \phi^{(0)}_{8}(\xi_7,{\bf r})
\rangle_{\xi_7}
\label{overlap}
\end{equation}
is the overlap function
between the ground state wave functions
of $^7$Be and $^8$B.
As one sees from Eq.~(\ref{Tap}), $\varphi^{(0)}({\bf r})$
plays an essential role in the $T$-matrix element.
Its calculation, however, requires
the wave functions of the two nuclei that are not readily available.
We, therefore, substitute it by a constant
${\mathfrak{S}}_{\rm exp}^{1/2}$ times a physically plausible
normalized function $ f^{(0)}({\bf r})$.
In this work we take it to be a solution of
\begin{equation}
(T_{r} + \tilde{V}_{\rm 17}({\bf r})- e_{17}) f^{(0)}({\bf r}) = 0,
\end{equation}
where $\tilde{V}_{\rm 17}({\bf r})$ is a Woods-Saxon potential
that is supposed to represent the single-particle potential of
$p$ in $^8$B.
Since the norm of $\varphi^{(0)}({\bf r})$
is the spectroscopic factor of $p$ in $^8$B,
${\mathfrak{S}}_{\rm exp}$ is its approximation
if $f^{(0)}({\bf r})$ is close to
$\varphi^{(0)}({\bf r})/||\varphi^{(0)} ( {\bf r})||^{1/2}$.
After the substitution of $f^{(0)}({\bf r})$,
the $T$-matrix element becomes
\begin{eqnarray}
{\mathfrak{T}}
&\approx& {\mathfrak{S}}_{\rm exp}^{1/2}
{\mathfrak{T}}^{\rm 3\mbox{-}body},
\nonumber \\
{\mathfrak{T}}^{\rm 3\mbox{-}body}&\equiv&
\langle \chi_{1}\chi_{7}|
U_{\rm A7}+ U_{\rm A1} + V_{\rm 17}|\Psi_{\rm 3\mbox{-}body}\rangle,
\label{T3}
\end{eqnarray}
where
\begin{equation}
\Psi_{\rm 3\mbox{-}body} =
\Omega^{(+)}
f^{(0)}({\bf r}) e^{i{\bf P} \cdot {\bf R}}
\label{Psi3}
\end{equation}
is the wave function of the total system with incident wave of $^8$B
consisting of a proton and an inert $^7$Be bound in a state
$f^{(0)}({\bf r})$ impinging on the target with momentum ${\bf P}$.

Thus, in the particular case under consideration, the exact $T$-matrix
element can approximately be calculated by three-body CDCC.
Furthermore, since the reaction is peripheral in $r$ as already mentioned,
the radial part of
$f^{(0)}({\bf r})$ in ${\mathfrak{T}}^{\rm 3\mbox{-}body}$
can be replaced by its asymptotic form
\begin{equation}
f_{\rm r}^{(0)}(r)  \sim \alpha W_{-\eta, 3/2}(2k_0 r) / r.
\label{alpha}
\end{equation}
Thus,
${\mathfrak{T}}$ is approximately proportional to
\begin{equation}
C = {\mathfrak{S}}_{\rm exp}^{1/2}\alpha.
\label{C}
\end{equation}
From the definitions of
${\mathfrak{S}}_{\rm exp}^{1/2}$ and $\alpha$, it is clear
that $C$ is in good approximation
the ratio of
$\varphi_{\rm r}^{(0)}(r)$ to $W_{-\eta, 3/2}(2k_0 r)/r$,
i.e. the ANC of $\varphi_{\rm r}^{(0)} (r)$ we look for.
Thus, one sees that the cross section is indeed proportional to $C^2$.
It should be noted that this proportionality holds only when the reaction
is peripheral in $r$, which is indeed the case
for the $^8$B breakup concerned.

In practice, we calculate the cross section
$\sigma_{\rm{calc}}$ from
${\mathfrak{T}}^{\rm 3\mbox{-}body}$
for each choice of $f^{(0)}({\bf r})$
with the three-body model CDCC as in the previous
work~\cite{CDCC-review1,CDCC-review2}.
We then obtain ${\mathfrak{S}}_{\rm exp}$ by taking the ratio of the
measured cross section to $\sigma_{\rm{calc}}$.
Since the amplitude $\alpha$ is readily known,
the ANC, $C$, can then be calculated by Eq.~(\ref{C}).
It should be noted that, since $f^{(0)}({\bf r})$ and
$\varphi^{(0)}({\bf r})$ may have quite different forms in the
inner region,
${\mathfrak{S}}_{\rm{exp}}^{1/2}$
may be quite different from the true spectroscopic amplitude,
$||\varphi^{(0)}({\bf r})||^{1/2}$, and
depend strongly on the choice of $f^0({\bf r})$.
In contrast, $C$ given by Eq.~(\ref{C})
is independent of $f^{(0)}({\bf r})$ and exactly the ANC of
$\varphi^{(0)}({\bf r})$ if the $^8$B breakup reaction is
genuinely peripheral with no contribution of the inner part of
the $^8$B wave function.
Once $C$ is well determined from a peripheral reaction, $S_{17}(0)$ can be
obtained by~\cite{Baye}
\begin{equation}
S_{17}(0) \approx 38.0(1-0.0013\bar{a}_{\rm s})C^2,
\label{S170}\end{equation}
where $\bar{a}_{\rm s}$ is the s-wave $p$-$^7$Be scattering
length that can be calculated by the $p$-$^7$Be interaction potential
used in the analysis.
The accuracy of the $S_{17}(0)$ extracted with the ANC method can
quantitatively evaluated by the dependence of $C$ on the choice of
$f^{(0)}({\bf r})$, as shown in Sec.~\ref{errorS17}.

Thus, we see that $S_{17}(0)$ can be extracted from the CDCC analysis
of $^8$B breakup based on the $p$+$^7$Be+A three-body model,
if one concentrates on the
$^8$B breakup into forward angles, i.e. $\theta_8 \la 4^\circ$.

\subsection{E-CDCC and hybrid scattering amplitude}
\label{E-CDCC}

In this subsection we describe the formalism of E-CDCC and how to
construct a hybrid scattering amplitude from the results of E-CDCC and
fully quantum-mechanical CDCC. The following formulation contains
two important extensions of E-CDCC. One is the inclusion of the intrinsic
spins of the projectile and its constituent fragments
and the other is a more accurate treatment of the Coulomb wave function
in the coupled-channel equations than in the previous formulation of
E-CDCC~\cite{Ogata2}. Detailed formalism of CDCC and its theoretical
foundations are given elsewhere~\cite{CDCC-review1,CDCC-review2,Austern}.

As already mentioned in Sec.~\ref{ANC},
the total wave function $\Psi_{\rm 3\mbox{-}body}$ of the
$p$+$^7$Be+A three-body system
satisfies the three-body Schr\"{o}dinger equation,
\begin{equation}
(H_{\rm 3\mbox{-}body}-E_{\rm in}-e_{17})
\Psi_{\rm 3\mbox{-}body}({\bf R},{\bf r})=0.
\label{sch}
\end{equation}
In this model, the $^8$B nucleus is assumed to consist of a $^7$Be
core in its ground state and a proton moving around it.
Its intrinsic wave function, therefore,
has the form $\Phi_{i,\ell S I m}({\bf r})$ where $\ell$ is
the orbital angular momentum of relative motion between $p$ and $^7$Be,
$S$ is the channel spin, and $I$ and $m$ are, respectively,
the total angular momentum
of $^8$B and its $z$-component.
The $z$-axis is taken to be parallel to the incident beam.
The index $i$ stands for the $i$ th of the states with the
quantum numbers \{$\ell$, $S$, $I$, $m$\}.
For simplicity we denote below the five channel-indices
\{$i$, $\ell$, $S$, $I$, $m$\} as $c$.
The set $\{\Phi_{c}\}$ consists of bound-states and
{\lq\lq}discretized-continuum-states''
as in Refs.~\cite{CDCC-review1,CDCC-review2}.
For the calculation of the latter,
the average method~\cite{Yahiro1,CDCC-review1,CDCC-review2,YK1}
and the pseudo-state methods~\cite{CDCC-review1,Matsumoto,PS1,PS2} are
widely used.
In the present work, we assume that the
$\Phi_{c}$ satisfy
\begin{equation}
\langle \Phi_{c'}({\bf r})|h |\Phi_{c}({\bf r})
\rangle_{\bf r}
=\epsilon_{i,\ell S I} \delta_{c'c},
\label{h}
\end{equation}
where $h$ is the internal Hamiltonian of $^8$B given by
$h= T_{r} +  \tilde{V}_{\rm 17}({\bf r})$,
and
$\epsilon_{i,\ell S I}$ is the intrinsic energy corresponding
to the channel \{$i$, $\ell$, $S$, $I$\}.
We take the following scheme of angular momentum coupling for
$\Phi_{c}({\bf r})$
\begin{eqnarray}
\Phi_{c}({\bf r})
&=&
\varphi_{i,\ell S I}(r)
\sum_{m_\ell,m_S}
(\ell m_\ell S m_S | I m)
i^\ell Y_{\ell m_\ell}(\hat{\bf r}) \zeta_{S m_S}
\nonumber \\
&\equiv&
\varphi_{i,\ell S I}(r)
\left[i^\ell Y_\ell(\hat{\bf r}) \otimes \zeta_S\right]_{Im},
\end{eqnarray}
where $Y_{\ell m_\ell}$ is the spherical harmonics and
$\zeta_{Sm_S}$ is the spin wave
function of $^8$B with channel spin $S$.

We expand $\Psi_{\rm 3\mbox{-}body}$ in terms
of the $\Phi_{c}({\bf r})$
as
\begin{equation}
\Psi_{\rm 3\mbox{-}body}({\bf R},{\bf r})
=
\sum_{c}
\Phi_{c}({\bf r})
e^{-i(m-m_0)\phi_R}
\chi_{c}(R,\theta_R),
\label{psinew2}
\end{equation}
where
the coefficient
$\chi_{c} (R,\theta_R) \exp[-i (m-m_0) \phi_R]$
of the expansion describes the c.m. motion of
the projectile relative to A in
channel $c$,
and $m_0$ is $m$ in the initial state.
Since the $\Phi_{c}$ are chosen to form an approximate complete set
in a finite region of space that is important
for the reaction concerned~\cite{Matsumoto},
the expansion of Eq.~(\ref{psinew2}) is accurate in that region of space.
Unknown coefficients $\chi_{c}$ of the expansion
are obtained by solving coupled-channel equations derived below.

Multiplying Eq.~(\ref{sch}) by $\Phi_{c'}^*({\bf r})$
from the left,
integrating over ${\bf r}$, and summing over
the spin states, one obtains the
coupled equations for the $\chi_{c}$
\begin{eqnarray}
e^{-i m'\phi_R}
({\cal T}_{R}+\epsilon_{i',\ell' S' I'}-E_{\rm in}-e_{17})
\chi_{c'}(R,\theta_R)
\nonumber \\
=
-\sum_{c}
F_{c'c}
e^{-i m \phi_R}
\chi_{c}(R,\theta_R),
\label{cceq}
\end{eqnarray}
where ${\cal T}_{R}$ is the reduced kinetic energy operator
defined by
\begin{equation}
{\cal T}_{R}
=
-\dfrac{\hbar^2}{2\mu}
\left[
\dfrac{1}{b}\dfrac{\partial}{\partial b}
\left(b\dfrac{\partial}{\partial b}\right)
+\dfrac{\partial^2}{\partial z^2}
\right]
+\dfrac{\hbar^2}{2\mu}
\dfrac{(m'-m_0)^2}{b^2}
\label{TR}
\end{equation}
in the cylindrical coordinates ${\bf R}= (z,b,\phi_R)$.
The impact parameter of the collision, $b$, is defined by
$b = \sqrt{x^2 +y^2}$ with ${\bf R} = (x,y,z)$
in the Cartesian coordinate.
The two arguments of $\chi_{c}$ in Eq.~(\ref{cceq})
are given by $R=\sqrt{z^2+b^2}$ and $\theta_R=\cos^{-1} (z/\sqrt{z^2+b^2})$.
In Eq.~(\ref{TR}) $\mu$ is the reduced mass of the $^8$B-A system.
In the peripheral region,
in which $b$ is large, we neglect
the last term of the r.h.s. of Eq.~(\ref{TR}),
which turns out to have no effect on numerical
results in the present case.
Thus,
\begin{equation}
{\cal T}_{R}
  \approx
-\dfrac{\hbar^2}{2\mu}
\left[
\dfrac{1}{b}\dfrac{\partial}{\partial b}
\left(b\dfrac{\partial}{\partial b}\right)
+\dfrac{\partial^2}{\partial z^2}
\right]
 \equiv
 {\cal T}'_R.
 \label{kinetic}
 \end{equation}
The form factor $F_{c'c}$ in Eq.~(\ref{cceq})
of the coupling between channels $c$ and $c'$ is
given by
\begin{eqnarray}
F_{c'c}({\bf R})
&=&
\left\langle
\Phi_{c'}
|
U_{\rm A7}+U_{\rm A1}
|
\Phi_{c}
\right\rangle_{\bf r}
\nonumber \\
&\equiv &
{\cal F}_{c'c}(R,\theta_R)
e^{-i(m'-m) \phi_R},
\label{FF}
\end{eqnarray}
where
we assume $U_{\rm A7}$ and $U_{\rm A1}$ to be central
potentials.
This assumption is valid as discussed in Ref.~\cite{Ogata2}
since we use E-CDCC only to obtain the scattering amplitude for large
values of $b$.
The multipole expansion of ${\cal F}_{c'c}$ is obtained from those
of the coupling potentials
\begin{eqnarray}
U_{\rm A7}(R_{\rm A7})
&=&
\sum_{\lambda}
u^{(\lambda)}_{\rm A7}(R, r/8)
\dfrac{4\pi}{\hat{\lambda}^2}
\sum_\nu
Y_{\lambda \nu}^*(\hat{\bf R})
Y_{\lambda \nu}(\hat{\bf r}),
\nonumber \\
U_{\rm A1}(R_{\rm A1})
&=&
\sum_{\lambda}
u^{(\lambda)}_{\rm A1}(R, 7r/8)
\dfrac{4\pi}{\hat{\lambda}^2}
\sum_\nu
Y_{\lambda \nu}^*(\hat{\bf R})
Y_{\lambda \nu}(\hat{\bf r}).
\nonumber
\end{eqnarray}
Then, ${\cal F}_{c'c}$ can be expanded
as
\begin{eqnarray}
{\cal F}_{c'c}(R,\theta_R)
&=&
\sum_{\lambda}
i^{\ell-\ell'}
\sqrt{4\pi}
(-)^{2I'+m-S}
\nonumber \\
& &
{}\times
\sqrt{\frac{(2\ell+1)(2\ell'+1)(2I+1)(2I'+1)}{(2\lambda+1)^3}}
\nonumber \\
& &
{}\times
(I -\!m\, I' m'| \lambda \;m'\!\!-\!m)
(\ell 0 \ell' 0|\lambda 0)
\nonumber \\
& &
{}\times
W(\ell \ell' II';\lambda S)\delta_{S'S}
\nonumber \\
& &
{}\times
C_{\lambda \,m'-m}
P_{\lambda \,m'-m}(\cos{\theta_R})
\nonumber \\
& &
{}\times
\int
\varphi_{i',\ell'S'I'}^*(r)
{\cal U}^{(\lambda)}(R,r)
\varphi_{i,\ell S I}(r)
r^2 dr,
\nonumber
\label{form2}
\end{eqnarray}
where $W(\ell \ell' II';\lambda S)$ is the Racah coefficient,
\begin{eqnarray}
C_{\lambda \nu}
\equiv
(-)^{(\nu+|\nu|)/2}
\sqrt{\dfrac{2\lambda+1}{4\pi}
\dfrac{(\lambda-|\nu|)!}{(\lambda+|\nu|)!}},
\nonumber
\end{eqnarray}
and ${\cal U}^{(\lambda)}(R,r) =
u^{(\lambda)}_{\rm A7}(R, r/8)+u^{(\lambda)}_{\rm A1}
(R, 7r/8)$.

Now we make the Coulomb-eikonal approximation~\cite{Kawai}
to $\chi_{c}$:
\begin{equation}
\chi_{c}(R,\theta_R)
\approx
\psi_{c}(b,z)
\phi_{c}^{{\rm C}(+)}(b,z),
\label{eikonal2}
\end{equation}
where $\phi_{c}^{{\rm C}(+)}$ is
the Coulomb wave function with outgoing scattered wave
and $\psi_{c}$ is the function to be determined
with the Eikonal approximation.
The Coulomb wave functions are given by
\begin{eqnarray}
\phi_{c}^{{\rm C}(+)}(b,z)
&=&
\dfrac{e^{-\pi\eta_c/2}}{(2\pi)^{3/2}}
\Gamma(1+i\eta_c)
e^{i{\bf K}_c \cdot {\bf R}}
F^c_{{\bf K}_c}({\bf R}),
\nonumber \\
F^c_{{\bf K}_c}({\bf R})
&\equiv&
G(-i\eta_c,1,i(K_c R- {\bf K}_c \cdot {\bf R}))
\nonumber
\label{Coulwf}
\end{eqnarray}
with $\Gamma$ the Gamma function and $G$ the confluent hypergeometric
function.
In actual calculation
we make use of the approximate asymptotic-form of $\phi_{c}^{{\rm C}(+)}$,
\begin{eqnarray}
\phi_{c}^{{\rm C}(+)}(b,z)
&\sim&
\dfrac{1}{(2\pi)^{3/2}}
\left(
1+\dfrac{\eta_c^2}{i(K_c R-{\bf K}_c\cdot {\bf R})}+...
\right)
\nonumber \\
& &
\times
e^{i({\bf K}_c \cdot {\bf R}+\eta_c\ln{(K_c R-{\bf K}_c\cdot {\bf R})})}
\nonumber \\
&\approx&
\dfrac{1}{(2\pi)^{3/2}}
e^{i(K_c z+\eta_c\ln{(K_c R-K_c z)})},
\label{Coulwf2}
\end{eqnarray}
which is valid for large values of $b$.
The boundary condition for $\psi_{c}$ in Eq.~(\ref{eikonal2}) is
$\displaystyle{\lim_{z\rightarrow -\infty}}\psi_{c}(b,z)=\delta_{c0}$,
where 0 denotes the incident channel,
so that $\chi_{c}$
satisfies the appropriate boundary condition
$\displaystyle{\lim_{z\rightarrow -\infty}}\chi_{c}(R,\theta_R)
=\delta_{c0}\phi_{c}^{{\rm C}(+)}(b,z)$.

We make the following two approximations
as in Ref.~\cite{Ogata2}.
(i) For sufficiently large $b$, the flux of $\phi_{c}^{{\rm C}(+)}$
is parallel to the $z$-axis.
(ii) Local semiclassical approximation~\cite{Watanabe}
to $\phi_{c}^{{\rm C}(+)}$,
\begin{equation}
\dfrac{\partial \phi_{c}^{{\rm C}(+)}(b,z)}{\partial z}
\approx
iK_c(R)\phi_{c}^{{\rm C}(+)}(R),
\label{LSCA}
\end{equation}
which is valid for a wave function in a slowly-varying potential.
The local momentum $K_c(R)$ is defined by
\begin{equation}
\dfrac{\hbar^2}{2\mu} K_c^2(R)
=
E_{\rm in}+e_{17}-\epsilon_{i,\ell S I}
-\dfrac{Z_{\rm 8}Z_{\rm A}e^2}{R},
\label{klcl}
\end{equation}
where $Z_8e$ ($Z_{\rm A}e$) represents the charge of $^8$B (A).
Further approximation to $\phi_{c}^{{\rm C}(+)}$ was made
in Ref.~\cite{Ogata2}, i.e. the $c$-dependence of $F^c_{{\bf K}_c}$
was neglected.
In the present paper, however, we
take account of that dependence using Eq.~(\ref{Coulwf2}).
This treatment of the Coulomb wave function
slightly improves the
agreement between the quantum-mechanical and eikonal
scattering-amplitudes for
large orbital angular momentum $L$ of $^8$B-A relative motion.
The explicit forms of the two amplitudes are given later
in Eqs.~(\ref{fC7})--(\ref{fCq}).

With Eq.~(\ref{kinetic}) and
the Coulomb-eikonal approximation, we obtain
\begin{eqnarray}
{\cal T}_{R} \chi_{c}^{J}(R,\theta_R)
&\approx&
{\cal T}'_{R} \left( \psi_{c}(b,z) \phi_{c}^{{\rm C}(+)}(b,z) \right)
\nonumber \\
&\approx&
\Bigg[
-\dfrac{i\hbar^2}{\mu}K_c(R)
\left(
\dfrac{\partial \psi_{c}(b,z)}{\partial z}
\right)
\phi_{c}^{{\rm C(+)}}(b,z)
\nonumber \\
& &
+
\left( {\cal T}'_{R} \phi_{c}^{{\rm C}(+)}(b,z) \right)
\psi_{c}(b,z)
\Bigg],
\nonumber \\
& &
\label{Tchi}
\end{eqnarray}
where the second-order derivative of $\psi_{c}$ is  neglected
since it is slowly varying compared with
$\phi_{c}^{{\rm C}(+)}$.
Inserting Eqs.~(\ref{FF}), (\ref{Coulwf2}), (\ref{Tchi}), and
\begin{eqnarray}
{\cal T}'_{R}\phi_{c}^{\rm C(+)}(b,z)
&=&
\left(
E_{\rm in}+e_{17}-\epsilon_{i,\ell S I}
-\dfrac{Z_{\rm 8}Z_{\rm A}e^2}{R}
\right)
\nonumber \\
& &
\times
\phi_{c}^{{\rm C}(+)}(b,z)
\nonumber
\end{eqnarray}
into Eq.~(\ref{cceq}), we arrive at
the following coupled-channel equations
\begin{eqnarray}
\dfrac{i\hbar^2}{\mu}
K_c(R)
\dfrac{d}{d z}\psi_{c}^{(b)}(z)
&=&
\sum_{c'}
{\mathfrak{F}}^{(b)}_{cc'}(z)
\;
{\cal R}^{(b)}_{cc'}(z)
\;
\psi_{c'}^{(b)}(z)
\nonumber \\
& &
\times
e^{i\left(K_{c'}-K_c \right) z},
\label{cceq4}
\end{eqnarray}
where the non-dynamical variable $b$ is put in a superscript,
the reduced coupling-potential ${\mathfrak{F}}_{cc'}$ is defined by
\begin{equation}
{\mathfrak{F}}^{(b)}_{cc'}(z)
=
{\cal F}^{(b)}_{cc'}(z)-
\dfrac{Z_{\rm 8}Z_{\rm A}e^2}{R}\delta_{cc'},
\end{equation}
and
\[
{\cal R}_{cc'}^{(b)}(z)
\equiv
\dfrac{
e^{i\eta_{c'}\ln{(K_{c'} R-K_{c'} z)}}
}
{
e^{i\eta_c\ln{(K_c R-K_c z)}}
}
=
\dfrac{
(K_{c'} R-K_{c'} z)^{i\eta_{c'}}
}
{
(K_c R-K_c z)^{i\eta_c}
}.
\]
Since Eq.~(\ref{cceq4}) is a set of first-order differential equations and
contains no coefficient with very large angular momentum $L$, one can
solve it with high computational speed and accuracy.

The eikonal scattering amplitude with Coulomb distortion is given by
$f_{c0}^{\rm E}=f_{c0}^{\rm Ruth}\delta_{c0}+f'^{\rm E}_{c0}$, where
$f_{c0}^{\rm Ruth}$ is the Rutherford amplitude and
\begin{eqnarray}
f'^{\rm E}_{c0}
&\equiv&
-\dfrac{(2\pi)^2\mu}{\hbar^2}
\int
\sum_{c'}
{\mathfrak{F}}^{(b)}_{cc'}(z)
\,
\phi_{c}^{*{\rm C}(-)}(b,z)
\phi_{c'}^{{\rm C}(+)}(b,z)
\nonumber \\
& &
\times
e^{-i(m-m_0)\phi_R}
\,
\psi_{c'}^{(b)}(z)
d{\bf R}
\label{f}
\end{eqnarray}
with
\begin{eqnarray}
\phi_{c}^{\rm C(-)}(b,z)
&=&
\dfrac{e^{-\pi\eta_c/2}}{(2\pi)^{3/2}}
\Gamma^*(1+i\eta_c)
e^{i{\bf K}'_c \cdot {\bf R}}
F^{c*}_{-{\bf K}'_c}({\bf R})
\nonumber \\
&\approx&
\dfrac{1}{(2\pi)^{3/2}}
e^{i{\bf K}'_c \cdot {\bf R}}
e^{-i\eta_c\ln{(K_c R+{\bf K}'_c \cdot {\bf R})})}.
\label{Coul-}
\end{eqnarray}
The outgoing c.m. momentum ${\bf K}'_c$ is chosen to be on the
$z$-$x$ plane, following the Madison convention.
For forward scattering in which
${\bf K}_{c}'$ is almost parallel to ${\bf K}_{c'}$,
i.e. to the $z$-axis,
we approximate, in the same way as in Ref.~\cite{Ogata2},
the phase-factor
$({\bf K}_{c'}-{\bf K}_{c}')\cdot {\bf R}$, which
appears in $\phi_{c}^{*{\rm C}(-)}(b,z) \phi_{c'}^{{\rm C}(+)}(b,z)$,
by
$-K_c\theta_f b\cos{\phi_R}+(K_{c'}-K_c)z$,
where $\theta_f$ is the scattering angle of $^8$B, and
$K_c R+{\bf K}'_c \cdot {\bf R}$ in
Eq.~(\ref{Coul-}) by $K_c R+K_c z$.
Under these approximations,
\begin{eqnarray}
f'^{\rm E}_{c0}
&\approx&
-\dfrac{\mu}{2\pi\hbar^2}
\int
e^{-i(m-m_0)\phi_R}
e^{-iK_c \sin{\theta_f}b\cos{\phi_R}}
\,
bdbd\phi_R
\nonumber \\
& &
\hspace*{-2mm}
\times{}
{\cal H}^{(b)}_c
\!\!\!
\int
\sum_{c'}
{\mathfrak{F}}^{(b)}_{cc'}(z)
\,
\psi_{c'}^{(b)}(z)
\,
e^{i (K_{c'}-K_c)z}
{\cal R}_{cc'}^{(b)}(z)
dz
\nonumber \\
\label{fC2}
\end{eqnarray}
with ${\cal H}^{(b)}_c \equiv \exp[2i\eta_c\ln{(K_c b)}]$.

The integration over $z$ in Eq.~(\ref{fC2}) can be done analytically
with the help of Eq.~(\ref{cceq4}), with the result
\begin{eqnarray}
{\cal I}
&\equiv&
\int
\sum_{c'}
{\mathfrak{F}}^{(b)}_{cc'}(z)
\,
\psi_{c'}^{(b)}(z)
\,
e^{i (K_{c'}-K_c)z}
{\cal R}_{cc'}^{(b)}(z)
dz
\nonumber \\
&=&
\dfrac{i\hbar^2}{\mu}
\int
K_c(R)
\left(
\dfrac{d}{d z}\psi_{c}^{(b)}(z)
\right)dz
\nonumber \\
&\approx&
\dfrac{i\hbar^2}{\mu}
\left[
K_c(R)
\psi_{c}^{(b)}(z)
\right]_{-\infty}^{\infty}
\nonumber \\
&\equiv&
\dfrac{i\hbar^2}{\mu}K_c
\left[
{\cal S}_{c0}^{(b)} - \delta_{c0}
\right],
\label{cceq6}
\end{eqnarray}
where $\partial K_c(R) / \partial z$ is neglected, since
$K_c(R)$, with large $b$, is a slowly-varying function of $z$.
The eikonal $S$-matrix elements are defined by
${\cal S}_{c0}^{(b)}=
\displaystyle{\lim_{z\rightarrow \infty}}\psi_{c}^{(b)}(z)$.
Insertion of Eq.~(\ref{cceq6}) into Eq.~(\ref{fC2}) leads to
\begin{eqnarray}
f'^{\rm E}_{c0}
&\approx&
\dfrac{K_c}{2\pi i}
\int
e^{-i(m-m_0)\phi_R}
e^{-iK_c \sin{\theta_f}b\cos{\phi_R}}
\,
bdbd\phi_R
\nonumber \\
& &
\times{}
{\cal H}^{(b)}_c
\left[
{\cal S}_{c0}^{(b)} - \delta_{c0}
\right].
\label{fC3}
\end{eqnarray}

We now transform the integration over $b$ to the summation over $L$
as in Ref.~\cite{Ogata2}. The resulting
eikonal scattering amplitude is
\begin{equation}
f^{\rm E}_{c0}
=
f_{c0}^{\rm Ruth}\delta_{c0}
+
\dfrac{2\pi}{i K_0}
\sum_L
f'^{\rm E}_{L;c0}\;
Y_{L \,m-m_0}(\hat{\bf K}'),
\label{fC7}
\end{equation}
where
\begin{equation}
f'^{\rm E}_{L;c0}
\equiv
\dfrac{K_0}{K_{c}}
{\cal H}^{(b_{c;L})}_c
\sqrt{\dfrac{2L+1}{4\pi}}
i^{(m-m_0)}
\left[
{\cal S}_{c0}^{(b_{c;L})}
-
\delta_{c0}
\right]
\label{fC8}
\end{equation}
with $K_c b_{c;L} = L + 1/2$.
The quantum mechanical scattering amplitude obtained by
CDCC is given by
\begin{equation}
f^{\rm Q}_{c0}
=
f_{c0}^{\rm Ruth}\delta_{c0}
+
\dfrac{2\pi}{i K_0}
\sum_L
f'^{\rm Q}_{L;c0}\;
Y_{L \,m-m_0}(\hat{\bf K}')
\label{fCq0}
\end{equation}
with
\begin{eqnarray}
f'^{\rm Q}_{L;c0}
&\equiv&
\sum_{J=|L-I|}^{L+I}
\sum_{L_0=|J-I_0|}^{J+I_0}
\sqrt{\dfrac{2L_0+1}{4\pi}}
\nonumber \\
& &
\times
(I_0 m_0 L_0 0 | J m_0)
(I m L \;m_0\!\!-\!m| J m_0)
\nonumber \\
& &
\times
(S_{i L \ell S I,i_0 L_0 \ell_0 S_0 I_0}^{J}-\delta_{i i_0}
\delta_{L L_0}\delta_{\ell \ell_0}\delta_{S S_0}\delta_{I I_0})
\nonumber \\
& &
\times
e^{i(\sigma_L+\sigma_{L_0})}
(-)^{m-m_0},
\label{fCq}
\end{eqnarray}
where $\sigma_L$ is the Coulomb phase shift
and $J$ is the total angular momentum of the three-body system.
We define the hybrid scattering amplitude by
\begin{eqnarray}
f^{\rm H}_{c0}
&=&
f_{c0}^{\rm Ruth}\delta_{c0}
+
\dfrac{2\pi}{i K_0}
\sum_{L < L_{\rm C}}
f'^{\rm Q}_{L;c0}\;
Y_{L \,m-m_0}(\hat{\bf K}')
\nonumber \\
& &
+
\dfrac{2\pi}{i K_0}
\sum_{L \ge L_{\rm C}}
f'^{\rm E}_{L;c0}\;
Y_{L \,m-m_0}(\hat{\bf K}')
\label{fh}
\end{eqnarray}
and use it in the analysis of $^8$B breakup.

The connecting-angular-momentum $L_{\rm C}$ between quantum mechanical
and eikonal amplitudes is chosen
so that $f_{L;c0}^{\rm E}$ agrees with $f_{L;c0}^{\rm Q}$ for
$L \ge L_{\rm C}$.
It should be noted that Eq.~(\ref{fh}) includes all quantum effects
and also the interference between amplitudes in the two $L$-regions.
This hybrid CDCC calculation turns out to
be capable of evaluating the nuclear and Coulomb breakup amplitudes with
very high accuracy and computational speed.
In fact, we have obtained the same accuracy
as standard fully quantum mechanical CDCC calculations
in about $1/10$ of computing time.

\section{Results and discussion}
\label{result}

\subsection{Numerical calculation}
\label{input}

For the wave functions of $^8$B that form the basis of the
CDCC expansion of the three-body system we make the following
assumptions. We assume that the ground state wave function of
$^8$B, with the $z$-component $m_0$ of the total spin $I_0=2$ of it,
is given by
\begin{eqnarray}
f_{m_0}^{(0)}({\bf r})
&=&
\{
w_1
\left[i Y_1(\hat{\bf r}) \otimes
\zeta_1
\right]_{2m_0}
+
w_2
\left[i Y_1(\hat{\bf r}) \otimes
\zeta_2
\right]_{2m_0}
\}
\nonumber \\
& &
\times
f_{\rm r}^{(0)}(r)
\nonumber \\
&\equiv&\;
w_1 f_{1,m_0}^{(0)}({\bf r})+
w_2 f_{2,m_0}^{(0)}({\bf r})
\label{f0}
\end{eqnarray}
in the channel spin representation,
neglecting the $\ell \neq 1$ components.
The values of the coefficients $w_1 \sim0.397$ and
$w_2 \sim0.918$ are obtained from
the result of the microscopic calculation~\cite{Akram} of $^8$B,
$C^2_{1/2} / C^2_{3/2}=0.157$,
where $C_{1/2}$ ($C_{3/2}$) is the ANC in the proton spin-orbit
coupling representation corresponding to
the $j\equiv|\vec{\ell} + \vec{\frac{1}{2}}|=1/2$ (3/2) state.
We further assume the effective $p$-$^7$Be interaction
for the ground state of $^8$B, $\tilde{V}_{17}^{(0)}$, to be independent
of $S$, hence the same radial dependence $f_{\rm r}^{(0)}(r)$ of
the $S=1$ and $S=2$ components in Eq.~(\ref{f0}).
This assumption is valid
since only the tail region, $r > R_{\rm N}$, of $f^{(0)}({\bf r})$
contributes to the $^8$B breakup process concerned.
The wave function of the $p$-$^7$Be scattering state
with outgoing scattered waves is given by
\begin{equation}
f_{S,m_{S}}^{(+)}({\bf k},{\bf r})
= \frac{i\varepsilon}{\varepsilon_{17} - T_r - V_{17} +i\varepsilon}
e^{i{\bf k} \cdot {\bf r}}\zeta_{S,m_{S}},
\hspace{2mm}
S=1\mbox{ or }2,
\label{scat8B}
\end{equation}
where
$e^{i{\bf k} \cdot {\bf r}}\zeta_{S,m_{S}}$ is the incident wave
with definite channel spin $S$ and its $z$-component $m_S$,
$\varepsilon_{17}$ is the incident energy in the c.m. frame of
the $p$-$^7$Be system,
and ${\bf k}$ is the incident
momentum.
The wave function with incoming scattered waves
$f_{S,m_{S}}^{(-)}({\bf k},{\bf r})$ corresponding to
$f_{S,m_{S}}^{(+)}({\bf k},{\bf r})$
used below is defined by
$f_{S,m_{S}}^{(-)}({\bf k},{\bf r})
=
f_{S,-m_{S}}^{(+)*}(-{\bf k},{\bf r})$.

We assume that $V_{17}$ has no tensor force component. We consider
the breakup of $^8$B to its scattering states at low $\varepsilon_{17}$,
and truncate the modelspace by $\varepsilon_{17} \le 10$ MeV.
In the experiment we analyze, however,
the cross section is measured for $\varepsilon_{17} $ in the range
of 500 keV $\le \varepsilon_{17} \le 750$ keV,
and it
turns out
that only coupling of those states with
$\varepsilon_{17}$ much lower than 10MeV need be taken into account
as described
in Sec.~\ref{resS17},
for which
only s-waves are affected by the nuclear part of $V_{17}$.
All the higher partial waves have negligible amplitude inside its
range because of the Coulomb and centrifugal barriers and there
have no coupling with the s-waves. Therefore, all the angular
momenta $\ell$, $S$ and $I$ are conserved in
$f_{S,m_{S}}^{(\pm)}({\bf k},{\bf r})$.
The channel spin $S$
is conserved also in s-waves since
$S = I$ that is a good quantum number.
Thus, one sees that
$f_{S,m_{S}}^{(+)}({\bf k},{\bf r})$
is the pure state of $S$, while $\ell$ and $I$ are mixed as
in the incident wave $e^{i{\bf k} \cdot {\bf r}}$.
It should be noted, however,
that $V_{17 }$ for the s-waves depends on $S$ as is evident from the large
$S$-dependence of the s-wave $p$-$^7$Be scattering lengths:
$a_{\rm s}^{S=2}=-7 \pm 3$ fm and
$a_{\rm s}^{S=1}=25\pm 9$ fm~\cite{Angulo}.

Furthermore, it turns out that the spin-dependent part of
the $p$-$^{208}$Pb optical potential has no effect on
the resulting elastic cross section at forward angles.
Thus, the channel spin $S$ is conserved during the
breakup process of $^8$B by the $^{208}$Pb target~\cite{CDCC-review1}.
Under these circumstances,
the $S$-dependent 3-body $T$-matrix element
in actual CDCC calculation,
${\mathfrak{T}}_{S,m_S m_0}^{\rm 3\mbox{-}body}$,
is given by
\begin{eqnarray}
{\mathfrak{T}}_{S,m_S m_0}^{\rm 3\mbox{-}body}
&=&
w_S
\langle
f_{S m_S}^{(-)}({\bf k},{\bf r})
e^{i{\bf P}' \cdot {\bf R}}
|
U_{\rm A7}+ U_{\rm A1}
\nonumber \\
& &
\times
|
\Omega^{(+)}
f_{S,m_0}^{(0)}({\bf r}) e^{i{\bf P} \cdot {\bf R}}
\rangle,
\nonumber \\
&\equiv&
w_S {\mathfrak{T}}_{S,m_S m_0}^{'{\rm 3\mbox{-}body}},
\end{eqnarray}
where
$\Omega^{(+)}$ is the wave matrix given in Eq.~(\ref{wavematrix}).

The unpolarized $^8$B breakup cross section calculated by CDCC, $\sigma$,
is given by
\begin{eqnarray}
\sigma
&\!\!=\!\!&
\frac{{\cal D} \rho}{2I_0+1}
\sum_{m_0 m_1 m_7}
\Big|\sum_{S m_S}
(\textstyle\frac{1}{2} m_1 \textstyle\frac{3}{2} m_7 | S m_S)
w_S
{\mathfrak{T}}_{S,m_S m_0}^{'{\rm 3\mbox{-}body}}
\Big|^2
\nonumber \\
&\!\!=\!\!&
\frac{{\cal D} \rho}{2I_0+1}
\sum_{S m_S m_0} w_S^2
\Big|
{\mathfrak{T}}_{S,m_S m_0}^{'{\rm 3\mbox{-}body}}
\Big|^2
\equiv
\sum_S
w_S^2 \sigma_S,
\label{sigcal}
\end{eqnarray}
where $m_1$ ($m_7$) is the $z$-component of the spin of $p$ ($^7$Be)
and the orthonormality of the Clebsch-Gordan coefficient is used.
We compare $\sigma$ with the unpolarized cross sections measured by
RIKEN experiment~\cite{RIKEN1,RIKEN2,RIKEN3}.
In the calculation of $\sigma_{1}$ and $\sigma_{2}$,
we use an $I$-independent $p$-$^7$Be effective
interaction $\tilde{V}_{17}$
as mentioned above. Then, the states of $^8$B
with fixed $S$ and $\ell$ and different values of $I$ are degenerate,
which greatly simplifies numerical calculation.
The value of $\bar{a}_{\rm s}$ in Eq.~(\ref{S170}) corresponding to
the spin-dependent ANC analysis is~\cite{Baye}
$\bar{a}_{\rm s}=w_1^2 a_{\rm s}^{S=1} + w_2^2 a_{\rm s}^{S=2}
\sim -2.8$ fm.

For $\tilde{V}_{17}^{(0)}$,
the single-particle potential of Esbensen and Bertsch (EB)~\cite{EB}
is used, except that
we neglect the spin-orbit
part of the potential and adjust the depth of the central potential to
reproduce the separation energy of $p$, 137 keV.
The scattering states for the p-waves are also calculated with
this $\tilde{V}_{17}^{(0)}$.
For d- and f-states with both $S=1$ and $S=2$,
we use the potential of Barker~\cite{Barker}.
For the s-state in the $S=2$ channel, we use the potential
of Barker that gives the s-wave $p$-$^7$Be
scattering length $a_{\rm s}^{S=2}$
of $-8$ fm.
For the s-state with $S=1$, we change the depth of the
potential of Barker to 25.7 MeV so that the resulting scattering
length $a_{\rm s}^{S=1}$ is 25 fm.

As for the distorting potential between $p$ ($^7$Be) and $^{208}$Pb
we adopt the global optical potential by Koning and Delaroche~\cite{KD}
(Cook~\cite{Cook}). We neglect the spin-orbit parts of the $p$-$^{208}$Pb
potential.
The multipoles for nuclear and Coulomb coupling-potentials
are included up to $\lambda=6$.
The discretized-continuum states of $^8$B are constructed by the average
method~\cite{Yahiro1,CDCC-review1,CDCC-review2,YK1}. The maximum
excitation energy of $^8$B is 10 MeV and 10, 20, 10, and 5
discretized-continuum states are taken for the $\ell=1$, 0, 2, and 3 states,
respectively. The resulting number of scattering-channels is 138.
The maximum values of $r$, $R$, and $L$ are, respectively,
200 fm, 1000 fm, and 12000.
The modelspace described above
turns out to give good convergence of
the resulting breakup cross sections.

\subsection{Analysis of $^8$B breakup experiment}
\label{resS17}

%
\begin{figure}[bp]
\begin{center}
 \includegraphics[width=0.45\textwidth,clip]{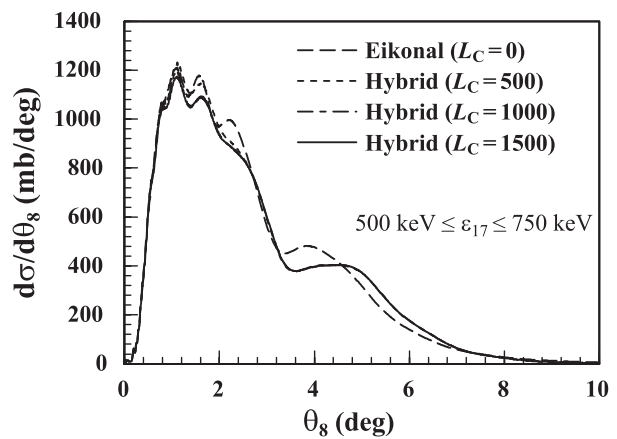}
 \caption{The $^8$B breakup cross section
 integrated over $\varepsilon_{17}$ in the range of 500--750 keV,
 as a function of $\theta_8$.
 The dotted, dash-dotted, and solid lines represent the results of the
 hybrid calculation of CDCC and E-CDCC with $L_{\rm C}=500$, 1000,
 and 1500, respectively. The result of E-CDCC, i.e. $L_{\rm C}=0$,
 is also shown by the dashed line.
 }
\end{center}
\end{figure}
Figure 2 shows the convergence of the hybrid calculation with
CDCC and E-CDCC for the breakup cross section of $^8$B by $^{208}$Pb
at 52 MeV/nucleon.
The  cross section integrated over
the excitation energy of $^8$B measured from the $p$+$^7$Be
threshold energy, $\varepsilon_{17}$,
from 500 keV to 750 keV
is plotted as a function of $\theta_8$.
The dashed line is the result of the E-CDCC calculation
($L_{\rm C}=0$) and the dotted, dash-dotted, and solid
lines correspond to the hybrid calculation with $L_{\rm C}=500$,
1000, and 1500, respectively. The figure shows that the hybrid
calculation converges with $L_{\rm C}=1000$. The difference
between the dash-dotted and solid lines is only less than about 1\%
in magnitude.
Thus, we regard the hybrid calculation of CDCC and E-CDCC with
$L_{\rm C}=1000$ as the fully quantum-mechanical CDCC calculation.

%
\begin{figure}[bp]
\begin{center}
 \includegraphics[width=0.45\textwidth,clip]{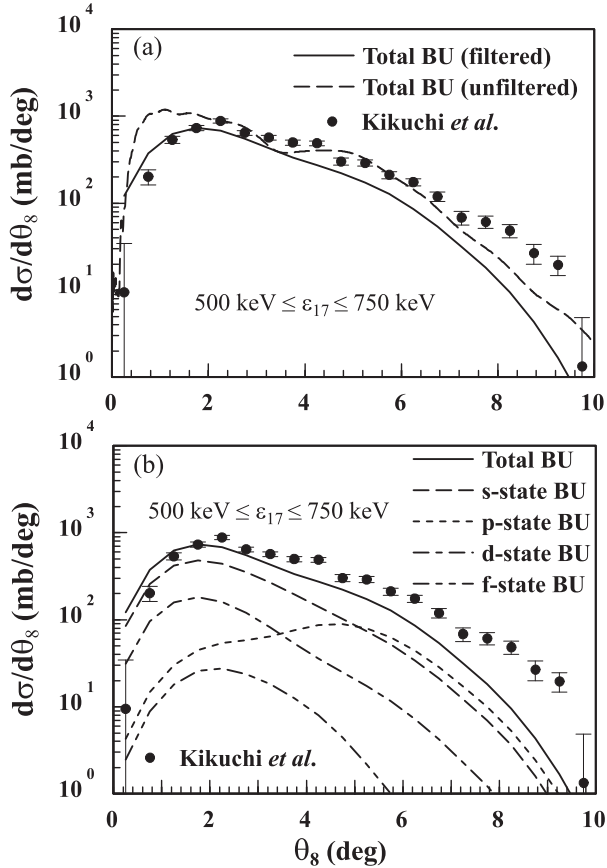}
 \caption{
 (a) The calculated $^8$B breakup cross sections with (solid line)
 and without (dashed line) the resolution and
 efficiency of the measurement~\protect\cite{Motobayashi}
 taken into account.
 The experimental data are taken from Ref.~\protect\cite{RIKEN2}.
 (b) Decomposition of the smeared $^8$B breakup cross section into
 the s- (dashed line), p- (dotted line), d- (dash-dotted line),
 and f-state (dash-two-dotted line) breakup components.
 }
\end{center}
\end{figure}
One sees in Fig.~2 the oscillation of the cross section at forward
angles around 2$^\circ$, which turns out to be due to the
interference between nuclear and Coulomb breakup-amplitudes.
Therefore, it can be concluded that nuclear interactions affect
the breakup cross section even at very forward angles.
However, as one sees below, this oscillation is not
observed in actual experimental data because of the limit of
the resolution of $\theta_8$ in the measurement.

Figure 3 shows the comparison between CDCC calculation and the
experimental data for the $^8$B breakup.
In order to take account of the experimental resolution and efficiency,
the theoretical result has been smeared out
by using the filtering table~\cite{Motobayashi} provided by
the authors of Ref.~\cite{RIKEN2}.
The upper panel shows the filtering effect on the cross section
calculated with CDCC,
where the solid and dashed lines represent the results
obtained with and without the smearing procedure.
In the lower panel, the smeared total breakup cross section
is decomposed into the s- (dashed line), p- (dotted line),
d- (dash-dotted line), and f-state (dash-two-dotted line)
breakup components of $^8$B.

One sees the smeared result of CDCC well agrees with the experimental
data at forward angles ($\theta_8 \la 4^\circ$), while it
underestimates the data at backward angles.
The lower panel shows the importance of the p-state breakup
for $\theta_8 \ga 5^\circ$, which implies that nuclear breakup
is significant in this angular region.
Thus, careful description of the p-state $^8$B wave function,
the resonance structures of it in particular, will be necessary to
reproduce the experimental data at backward angles.
The underestimation there may be due also to dynamical roles played by the
excited-core component in $^8$B,
$^7$Be(1/2$^-$)$\otimes p$(3/2$^-$),
that is neglected in the present work as discussed in Sec.~\ref{ANC}.
Although further study in this line will be very
interesting, it is beyond the scope of the present paper.
Additionally, the filtering table used was made with assuming the
s-state breakup of $^8$B. Thus, quantitative comparison between the
calculation and the experimental data, to extract $S_{17}(0)$ with
high accuracy, can only be done in the region where the
s-state breakup cross section is dominant.
Therefore, below we use the data for $\theta_8 \le 4^\circ$ to determine
${\mathfrak{S}}_{\rm{exp}}$ by
\begin{equation}
{\mathfrak{S}}_{\rm{exp}}
=
\frac{\sigma_{\rm{exp}}}{\sigma},
\label{Spec}
\end{equation}
where $\sigma$ is given by Eq.~(\ref{sigcal}) and
$\sigma_{\rm{exp}}$ is the corresponding experimental data.

%
\begin{figure}[htbp]
\begin{center}
 \includegraphics[width=0.45\textwidth,clip]{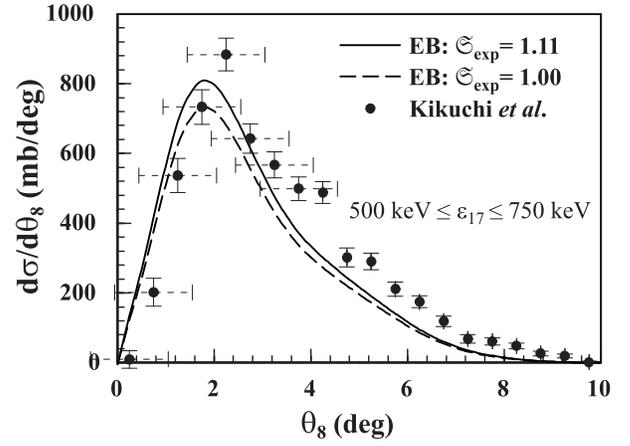}
 \caption{
 The result of the $\chi^2$-fit of the breakup cross section
 calculated with the $^8$B single-particle potential of
 Esbensen and Bertsch (EB)~\protect\cite{EB}.
 The solid line is the same as that
 in each panel in Fig.~3 but multiplied by the spectroscopic factor
 ${\mathfrak{S}}_{\rm exp}=1.11$.
 The result with ${\mathfrak{S}}_{\rm exp}=1.00$ is also shown by the
 dashed line for comparison.
 For the eight data points below $4^\circ$, which are used in the fitting
 procedure, we have put the horizontal bar that represents the
 resolution of $\theta_8$~\protect\cite{RIKEN2} of the measurement.
 }
\end{center}
\end{figure}
Figure 4 shows the result of the $\chi^2$-fit of the theoretical
calculation to the experimental data. The solid line is
the same as that in each panel in Fig.~3 but multiplied by
the spectroscopic factor ${\mathfrak{S}}_{\rm exp}=1.11$.
The dashed line represents the result before being fitted, i.e.
with ${\mathfrak{S}}_{\rm exp}=1.00$.
Each horizontal bar put on the eight data points below $4^\circ$
does not represent a statistical error but it
shows the range of $\theta_8$ in which the breakup cross sections
contribute to each data point~\protect\cite{RIKEN2}.
The value of $\chi^2$ per datum
obtained is 8.5. Although the quality of the fit is
not good, it should be noted that estimation of error-propagation
in the present case, i.e. comparison between smeared experimental data
and a smeared numerical result, is very complicated; the value of
$\chi^2$ per datum shown above takes account of no error with respect
to $\theta_8$.
The value of $\alpha$ in Eq.~(\ref{alpha}) for the EB $^8$B wave-function
is 0.704 fm$^{-1/2}$.
With the values of ${\mathfrak{S}}_{\rm exp}$ and $\alpha$,
the ANC $C$ is obtained from Eq.~(\ref{C}) as $C=0.740$ fm$^{-1/2}$.
One can then determine $S_{17}(0)$ by inserting
this result into Eq.~(\ref{S170}), together with
$\bar{a}_{\rm s}=-2.8$ fm, i.e.
$S_{17}(0)=20.9$ eV b.
In the next subsection we quantitatively evaluate uncertainties of
the extracted $S_{17}(0)$.

\subsection{Uncertainties of the extracted $S_{17}(0)$}
\label{errorS17}

First, we evaluate the uncertainty of $S_{17}(0)$ that comes from the
use of the ANC method, just in the same way as in Refs.~\cite{Ogata,OgataND}.
We show in Fig.~5 the result of CDCC calculation with
the p-state single-particle potential of $^8$B by Kim {\it et al.}~\cite{Kim}
(the dashed line) and that multiplied by ${\mathfrak{S}}_{\rm exp}=0.867$
(the solid line); the latter agrees with the result shown
in Fig.~4 within 1\%.
One sees that ${\mathfrak{S}}_{\rm exp}$ indeed depends on the p-state
potential.
%
\begin{figure}[htbp]
\begin{center}
 \includegraphics[width=0.45\textwidth,clip]{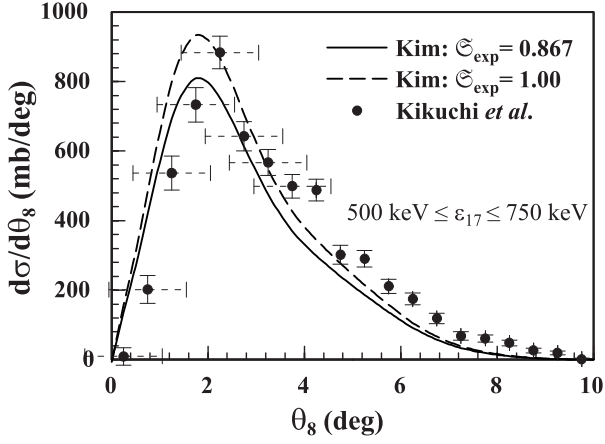}
 \caption{
 Same as in Fig.~4 but for the CDCC calculation with Kim's single-particle
 potential~\protect\cite{Kim} for the p-state of $^8$B; the spectroscopic
 factor ${\mathfrak{S}}_{\rm exp}$ obtained
 is 0.867.
 The result with Kim's potential before $\chi^2$-fitting, namely with
 ${\mathfrak{S}}_{\rm exp}=1.00$, is also shown by the dashed line for
 comparison.
 }
\end{center}
\end{figure}
Actually,
$\alpha$ in Eq.~(\ref{alpha})
for Kim's potential is 0.796 fm$^{-1/2}$,
which is quite different from the value for the EB potential, 0.704
fm$^{-1/2}$,
because of
the difference in the geometry of the
two potentials of about 20\%.
On the contrary,
the value of $C$ in Eq.~(\ref{C}) calculated
with Kim's potential is
0.741 fm$^{-1/2}$ that agrees very well with the result obtained
with the EB potential shown in Sec.~\ref{resS17}.
This result shows that the ANC method works
with very high accuracy in the present analysis, i.e.
the error of the ANC method is negligible.

Second, we estimate the effect of ambiguity of the distorting potentials
used in the CDCC calculation on the obtained $S_{17}(0)$.
For this purpose we use alternative optical potentials for
$p$-$^{208}$Pb and $^7$Be-$^{208}$Pb.
For the $p$-$^{208}$Pb potential,
we modify the parameter set of Koning and Delaroche
so that the calculated elastic cross section without spin-orbit terms
reproduces the one obtained with the full components of the optical
potential. The parameter set thus obtained is
$V_V=39.82$ MeV, $W_V=0.776$ MeV, and $W_D=13.47$ MeV, with the same
notation as in Ref.~\cite{KD}; all other parameters are not changed.
For the $^7$Be-$^{208}$Pb potential,
we use a single-folding model with the
nucleon-$^{208}$Pb potential of Koning and Delaroche,
neglecting the spin-orbit terms.
The density distribution of $^7$Be
is assumed to be of the Gaussian form with the range that
reproduces the rms matter radius of
$^7$Be of 2.48 fm~\cite{Tanihata}.
The difference between the results calculated with the original and
alternative optical potentials is found to be negligibly small
(not shown).
We conclude, therefore, that the error of $S_{17}(0)$ that
comes from the ambiguity of the distorting potentials
in the present case is negligibly small.

%
\begin{figure}[htbp]
\begin{center}
 \includegraphics[width=0.45\textwidth,clip]{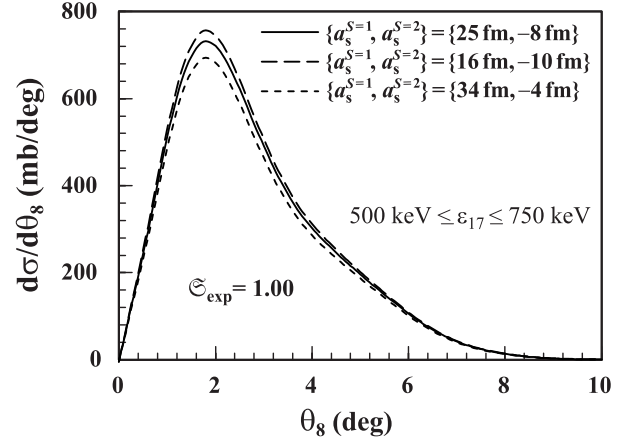}
 \caption{
 Dependence of the $^8$B breakup cross section on the s-state single-particle
 potentials of $^8$B for the $S=1$ and $S=2$ channels.
 The solid line represents the
 result with the choice of the potentials that
 corresponds to the s-wave $p$-$^7$Be scattering lengths
 $\{ a_{\rm s}^{S=1}, a_{\rm s}^{S=2} \}
 = \{ 25 \mbox{ fm}, -8 \mbox{ fm} \}$.
 The dashed (dotted) line shows the
 result with the s-state potentials corresponding to
 $\{ a_{\rm s}^{S=1}, a_{\rm s}^{S=2} \}
 = \{ 16 \mbox{ fm}, -10 \mbox{ fm} \}$
 ($\{ 34 \mbox{ fm}, -4 \mbox{ fm} \}$).
 }
\end{center}
\end{figure}
Finally, we estimate the uncertainty of $S_{17}(0)$ due to
the ambiguity of the s-state single-particle
potentials of $^8$B for the $S=1$ and $S=2$ channels
by changing the depth of the potential
for the $S=1$ ($S=2$) channel ranging from 23.3 MeV (52.8 MeV)
to 30.5 MeV (57.6 MeV), which gives the range of the
resulting s-wave $p$-$^7$Be
scattering length $a_{\rm s}^{S=1}$ ($a_{\rm s}^{S=2}$)
from $16$ fm ($-10$ fm) to $34$ fm ($-4$ fm);
these values of $a_{\rm s}^{S=1}$ and $a_{\rm s}^{S=2}$
are within the range of errors of
the experimental values of them~\cite{Angulo}.
In Fig.~6 the solid, dashed, and dotted
lines represent the results with the s-state potentials corresponding to
$\{ a_{\rm s}^{S=1}, a_{\rm s}^{S=2} \} = \{ 25 \mbox{ fm}, -8 \mbox{ fm} \}$,
$\{ 16\mbox{ fm},-10\mbox{ fm} \}$, and
$\{ 34\mbox{ fm},-4\mbox{ fm} \}$, respectively.
One sees the breakup cross section slightly depends on the choice
of the s-state potentials of $^8$B;
the resulting $S_{17}(0)$ is found to vary from 20.3 eV b to 21.9 eV b.
Thus, we evaluate the error of $S_{17}(0)$ to be $+4.9\%$/$-2.9\%$.

Summarizing the discussion given above, the result
of the present paper is as follows.
The central value of $S_{17}(0)$ is 20.9 eV b.
The theoretical error of the extracted $S_{17}(0)$ concerned with
the s-wave
$p$-$^7$Be scattering length is $+4.9\%$/$-2.9\%$.
After including the 8.4\% systematic experimental error,
we obtain
\[S_{17}(0)=20.9_{-0.6}^{+1.0}\mbox{ (theo)}\pm 1.8\mbox{ (expt) }
\;\; {\rm eV b}.
\]

\subsection{Discussion of the extracted $S_{17}(0)$}
\label{discussion}

The main result of the present paper is
$S_{17}(0)=20.9_{-0.6}^{+1.0}$ (theo) $\pm 1.8$ (expt) eV b
derived from $^{208}$Pb($^8$B, $p+^7$Be)$^{208}$Pb at 52 MeV/nucleon.
This value is significantly larger than
$S_{17}(0)=18.9 \pm 1.8$ eV b~\cite{RIKEN3}
obtained in the previous
analysis of the same experiment
with the first-order perturbation theory.
In order to clarify the reason for the difference, we discuss
roles of nuclear interaction, E2 transitions, and higher-order processes.
%
\begin{figure}[htbp]
\begin{center}
 \includegraphics[width=0.45\textwidth,clip]{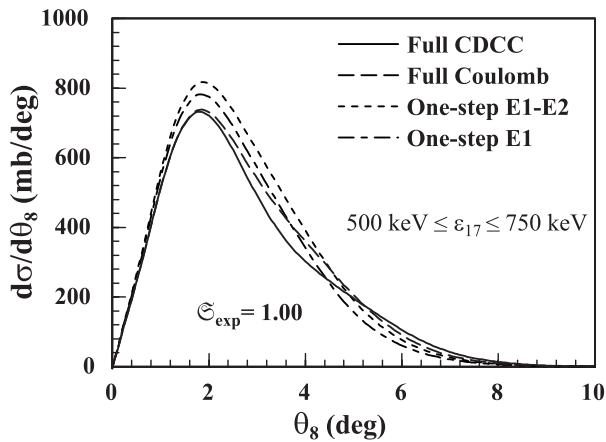}
 \caption{
 Same as in Fig.~3 with different assumptions for CDCC.
 The lines represent the results of CDCC with nuclear and
 Coulomb couplings and all higher-order processes (solid line),
 CDCC without nuclear breakup (dashed line),
 one-step CDCC with Coulomb E1 and E2 components (dotted-line),
 and one-step CDCC with Coulomb E1 only that corresponds to
 the first-order virtual photon absorption theory (dash-dotted line).
 }
\end{center}
\end{figure}
In Fig.~7 the solid line
is the same as
in the upper panel in Fig.~3
that shows the result of full CDCC.
The dashed line corresponds to the calculation without
breakup components of nuclear coupling potentials.
It should be noted that their diagonal components
are included in the calculation to take
account of the absorption of the flux
of the incident particle
by the target nucleus.
One sees that the dashed line agrees with the solid line
below $2^\circ$ but deviates from it for $\theta_8>2^\circ$.
As a result, the value of $S_{17}(0)$
obtained, $S_{17}(0)=20.0_{-1.8}^{+1.9}$ eV b,
is smaller than
$S_{17}(0)=20.9_{-0.6}^{+1.0}$ (theo) $\pm 1.8$ (expt) eV b
obtained with the full CDCC calculation mentioned above.

The dash-dotted line in Fig.~7 represents the result of first-order
iterative solutions of CDCC, designated as one-step CDCC,
without nuclear breakup, including only $\lambda=1$ (E1)
component of the Coulomb interaction;
this calculation is essentially the same as
the first-order perturbation theory
(virtual photon theory) used in the previous
analysis of the experimental data~\cite{RIKEN3}.
It overestimates the solid
line above 1.5$^\circ$ and the resulting value of $S_{17}(0)$ is
$19.0_{-1.7}^{+1.9}$ eV b,
which agrees well with the value $18.9 \pm 1.8$ eV b
obtained in Ref.~\cite{RIKEN3}.

If one includes the $\lambda=2$ (E2) component
in the one-step CDCC calculation of Coulomb breakup,
the dotted line in Fig.~7 is obtained;
the coupling potentials with $\lambda=2$ are artificially
multiplied by 0.7, following the analysis of the MSU data~\cite{MSU2}.
One sees from Fig.~7 that inclusion of the reduced
E2 component somewhat increases the breakup cross section, which
results in
further decrease of $S_{17}(0)$ to
$S_{17}(0)=17.9_{-1.6}^{+1.7}$ eV b.
This result is consistent with the conclusion of Ref.~\cite{MSU2},
in which $S_{17}(0)=17.8^{+1.4}_{-1.2}$ eV b was derived with
first-order perturbation theory including both E1 and the reduced
E2 components.
If the E2 component is not scaled,
the resulting value of $S_{17}(0)$
is 16.7 eV b.
This value is about 20\% less than the result of full CDCC,
20.9 eV b. The difference is due to the nuclear and
higher-order Coulomb processes since the coupling potentials
with $\lambda \ge 3$ are found to have little effect on the
total breakup-cross-section at $\theta_8 \le 4^\circ$.

Thus, description of $^8$B breakup process with
nuclear and Coulomb breakup processes of
both the E1 and E2 transitions and higher-order processes
is a key to solve a puzzle recognized so far
of
the discrepancy between the values of $S_{17}(0)$
extracted from direct $p$-capture reactions,
$S_{17}(0)=22.1 \pm 0.6$ (expt) $\pm 0.6$ (theo) eV b,
and indirect $^8$B dissociation experiments.
In order to clarify the role of these components in $^8$B
breakup reaction in  general,
it is necessary to carry out analyses to wider range of data,
such as those measured at MSU~\cite{MSU1,MSU2} and GSI~\cite{GSI1,GSI2},
including the data on other
quantities than that dealt with in the present work.
Analysis of parallel momentum ($p_{//}$) distribution of
$^7$Be-fragment after breakup of $^8$B is particularly important,
since the role of E2 component was determined from it~\cite{MSU1,MSU2}.
In the analysis in Ref.~\cite{MSU2} the result of first-order
perturbation theory
with an adjustable parameter was compared with CDCC calculation
that is essentially the same as in the present paper.
Because of the agreement of the two results, it was concluded
that higher order processes were unimportant
without quantitative evaluation of the nuclear and higher-order
Coulomb processes as done in the present work.
Actually, the contribution is about
20\% of the total as already described.
A CDCC analysis of the $p_{//}$ distribution has been done
by Mortimer {\it et al.}~\cite{Mortimer}, where the dependence of
the calculated result on the strength of
the Coulomb $\lambda=2$ (E2) coupling potentials
was discussed.
They found that the E2 potentials must be
multiplied by 1.6 to reproduce the experimental data,
although the origin of the enhancement was not clear.
Therefore, further investigation of
the accuracy of the enhancement factor 1.6 seems necessary.
Unfortunately, filtering table for the MSU experiment is not
available to the present authors.
At this stage, therefore, we can quantitatively extract
$S_{17}(0)$ with CDCC and the ANC method only from the $^8$B
breakup experiment done at RIKEN.
Nevertheless, comparison of our pure theoretical result with that
shown in Ref.~\cite{Mortimer} is very interesting,
which will be a subject of our future work.

\subsection{ANC analysis of the direct measurement}
\label{direct}

In this subsection we report on our application of
the ANC method to the $p$-capture reaction,
$^7$Be($p,\gamma$)$^8$B, at energies of 116--362 keV~\cite{Junghans2}.
We assume direct capture of $p$ by $^7$Be to the ground state of $^8$B
by a pure E1 transition. A simple potential-model
is assumed for the bound and the scattering states of the
$p$+$^7$Be system,
using the single-particle potential of Sec.~\ref{input}.
The s- and d-state components in the initial
scattering-state wave-function are taken into account,
and the intrinsic spins of $p$ and $^7$Be are included in the
channel spin representation described in Sec.~\ref{input}.
In Fig.~8 the solid line shows the calculated $S_{17}$ as a function
of
the incident energy $E_{\rm c.m.}$ of the reaction in the c.m. frame
of the $p$-$^7$Be system .
The plotted $S_{17}(E_{\rm c.m.})$
has been multiplied by the spectroscopic factor
${\mathfrak{S}}_{\rm exp}=1.15$ obtained by
the $\chi^2$-fit to the data below 362 keV.
%
\begin{figure}[htbp]
\begin{center}
 \includegraphics[width=0.43\textwidth,clip]{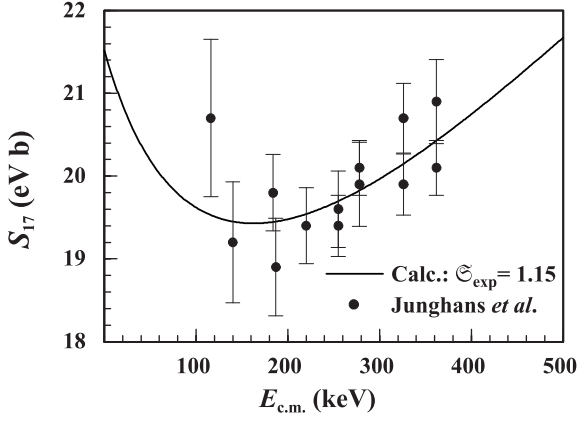}
 \caption{
 The astrophysical factor $S_{17}$ as a function of the incident energy
 $E_{\rm c.m.}$ of the reaction in the c.m. frame of the $p$-$^7$Be system.
 The solid line is the result of the calculation with a simple potential
 model, multiplied by ${\mathfrak{S}}_{\rm exp}=1.15$.
 The experimental data are taken from Ref.~\protect\cite{Junghans2}.
 }
\end{center}
\end{figure}
The calculated $S_{17}(E_{\rm c.m.})$ is not reliable at high energies,
say $E_{\rm c.m.}>400$ keV,
since the inner part of the $^8$B
wave function plays an important role.
At low energies, however,
the ANC method works very well and a reliable value of the ANC $C$,
and consequently $S_{17}(0)$, can be derived from the $\chi^2$-fitting
procedure.
The value of $S_{17}(0)$ thus obtained is 21.7 eV b. Uncertainties
of the $S_{17}(0)$
are evaluated in the same way as in Sec.~\ref{errorS17} and found
to be $+0.37$/$-0.24$ eV b, almost all of which is due to
the ambiguity of the s-state potentials for the $S=1$ and $S=2$ channels
of the $p$-$^7$Be system.
Including the 2.3\% systematic experimental error,
we obtain the following result:
$S_{17}(0)=21.7^{+0.37}_{-0.24}$ (theo) $\pm 0.50$ (expt) eV b.
This result is consistent with both the value extracted from
$^8$B dissociation,
$S_{17}(0)=20.9_{-0.6}^{+1.0}$ (theo) $\pm 1.8$ (expt) eV b,
described in the previous subsections,
and the result obtained in Ref.~\cite{Junghans2},
$S_{17}(0)=22.1 \pm 0.6$ (expt) $\pm 0.6$ (theo) eV b,
although no evaluation of the ambiguity due to the uncertainty of
the s-wave $p$-$^7$Be scattering lengths, as the one described
in Sec.~\ref{errorS17}, is made in Ref.~\cite{Junghans2}.

\section{Summary}
\label{summary}

The principal result of the present paper is
the value of $S_{17}(0)$ of
$20.9_{-0.6}^{+1.0}$ (theo) $\pm 1.8$ (expt) eV b
obtained by an analysis of the cross section of the
$^8$B breakup reaction $^{208}$Pb($^8$B, $p+^7$Be)$^{208}$Pb
at 52 MeV/nucleon measured at RIKEN~\cite{RIKEN1,RIKEN2,RIKEN3}
by means of the method of
continuum-discretized coupled-channels~\cite{CDCC-review1,CDCC-review2}
(CDCC) combined
with the asymptotic normalization coefficient (ANC) method~\cite{Akram}.
The value is consistent with the one extracted from the precise
measurement of the cross section of direct capture
$^7$Be($p,\gamma$)$^8$B,
$S_{17}(0)=22.1 \pm 0.6$ (expt) $\pm 0.6$ (theo) eV b~\cite{Junghans2}.

The CDCC calculation is based on the three-body model of $p$,
$^7$Be, and the target $^{208}$Pb nucleus, with $^7$Be staying
in the ground state throughout the reaction process. This model
is shown to be adequate in the breakup reaction in which the
fragments $p$ and $^7$Be are ejected in very forward angles, less
than 4$^\circ$, which is the case for the experimental data analyzed
in this work. Nuclear and Coulomb E1 and E2 transitions and multi-step
processes of all-order are included in the calculation. The calculated
cross section is shown to be proportional to the squared asymptotic
amplitude of the overlap of the ground state wave functions of $^7$Be
and $^8$B, which is a necessary condition for the applicability of the
ANC method.
The CDCC calculation is made efficient by the use of
the eikonal CDCC method~\cite{Ogata2} (E-CDCC)
for large angular momentum partial waves.
Improvements of E-CDCC are made to cope with the Coulomb distortions
of the incident and scattered waves of relative motion and intrinsic
spins of the projectile and the ejectiles.

The main source of the error in the obtained $S_{17}(0)$ is the ambiguity
of the  s-wave $p$-$^7$Be interaction potentials for the channel spin
$S = 1$ and $S = 2$ states.
The errors associated with the use of the ANC method and the ambiguity
of the distorting potentials are both negligible.
\begin{figure}[htbp]
\begin{center}
\includegraphics[width=0.48\textwidth,clip]{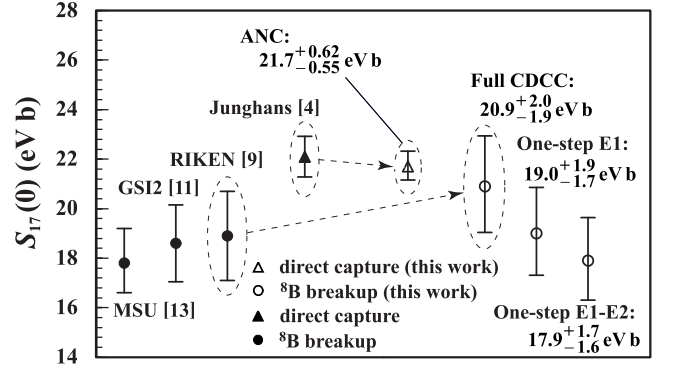}
 \caption{
 The values of $S_{17}(0)$ extracted by the ANC method
 are shown by open symbols: the circles are obtained from
 three-types of CDCC analysis of $^8$B breakup
 and the triangle is the result of analysis of $^7$Be($p,\gamma$)$^8$B
 with a simple potential model.
 The error of each result is obtained by adding the theoretical and
 experimental errors in quadrature.
 The results are compared with $S_{17}(0)$ extracted from
 the  precise  measurement of direct capture $(p,\gamma)$
 cross section~\protect\cite{Junghans2} (closed triangle)
 and those obtained from $^8$B dissociation with
 first-order perturbation theory (closed circles).
 }
\end{center}
\end{figure}
Calculations of $S_{17}(0)$  with some simplified assumptions are
summarized in Fig.~9.  The results of first-order iterative
solutions of CDCC, designated as one-step CDCC, correspond to and
agree with those of first-order perturbation theory in
the previous work~\cite{RIKEN1,RIKEN2,RIKEN3,MSU1,MSU2}.
The inclusion of the Coulomb
quadrupole (E2) transitions, scaled by 0.7, in the one-step calculation
decreases $S_{17}(0)$ by about 6\%, and multistep processes increase
it by about 20\%.
This shows the crucial importance of accurate
description of the $^8$B breakup process by CDCC including nuclear
and Coulomb E1 and E2 transitions and all higher-order processes.
It will be interesting to apply the method of the present paper to
analyses of wider range of experiments such as those in
Refs.~\cite{GSI1,GSI2,MSU1,MSU2}.
It may, however, be that the three-body CDCC used in the present
paper is not valid  in general and the use of CDCC with four-body
model such as the one in Ref.~\cite{Matsumoto2} becomes necessary.

\section*{Acknowledgments}
The authors would like to thank T. Motobayashi for helpful discussions
and providing detailed information on the experiment.
The authors also thank M. Kawai for fruitful discussions
and careful reading of the manuscript.
This work has been supported in part by the Grants-in-Aid for
Scientific Research of Monbukagakusyou of Japan.


\end{document}